# A Cache Management Scheme for Efficient Content Eviction and Replication in Cache Networks

Muhammad Bilal and Shin-Gak Kang

*Abstract*— To cope with the ongoing changing demands of the internet, 'in-network caching' has been presented as an application solution for two decades. With the advent of information-centric network (ICN) architecture, 'in-network caching' becomes a network level solution. Some unique features of ICNs, e.g., rapidly changing cache states, higher request arrival rates, smaller cache sizes, and other factors, impose diverse requirements on the content eviction policies. In particular, eviction policies should be fast and lightweight. In this study, we propose cache replication and eviction schemes, Conditional Leave Cope Evwhere (CLCE) and Least Frequent Recently Used (LFRU), which are well suited for the ICN type of cache networks (CNs). The CLCE replication scheme reduces the redundant caching of contents; hence improves the cache space utilization. LFRU approximates the Least Frequently Used (LFU) scheme coupled with the Least Recently Used (LRU) scheme and is practically implementable for rapidly changing cache networks like ICNs.

## I. INTRODUCTION

IN the recent past there has been a profound increase in the internet connectivity, and with the emergence of new internet applications, such as online social network applications, live video streaming, video sharing, multi user online gaming and IoT, internet semantics have changed from host centric to content centric. To satisfy the needs of emerging internet applications, the current internet architecture has adopted several application layer solutions known as Over-the-Top (OTT) applications, such as Content Delivery Network (CDN), web caching, and peer-to-peer networking [1-6]. In fact, the additions of new OTT applications are leading us towards very complex internet architecture. Van Jacobson identified a basic paradigm shift in internet services [7] and introduced a fresh concept of internet architecture, known as Information-Centric

Networking (ICN). In the ICN model, 'in-network caching' is an integral part of the ICN service framework [8-9]. Unlike CDNs, web caching and P2P networking, ICN is a network layer solution; hence, all ICN enabled routers are responsible for storing downloaded content for a limited time. In the ICN paradigm users request content by content name and the network performs mapping between the request and the location of the content across the network. The ICN routers create a cache network (CN), and the analysis of the CN is of prime importance in analyzing the ICN. The end user experience depends on how fast the requested content can be found and delivered; it is important to store the most popular content in cache nodes.

In-network caching is not only influenced by the ICN architecture and framework but it also depends on the various application specific demands, operator related control and business models. In this work, we design and propose a content eviction and replication scheme for an ICN network that achieves a high content hit rate. The goal of the content eviction scheme is to maintain the most popular content stored in the cache with the least processing complexity, while the goal of the replication scheme is to replicate the content in the caches where the local popularity of the content is high. We also argue that the application specific, operator related control and business model demands should be incorporated in only the replication scheme.

The content storage in a network cache is governed by two caching strategies: the replication algorithm and the content eviction algorithm. Replication algorithms are used to spread copies of content across the network. Several replication algorithms have been discussed in the literature. According to Leave Copy Everywhere (LCE) [7,10] the requested content is stored in all cache nodes (an ICN enabled router) on its way to the requesting user; hence, it imposes a huge burden on the entire network to run a replacement algorithm on each cache. In Leave a Copy Down (LCD) [11-12] the content is replicated to only downward cache nodes; this scheme requires a long time to replicate the content in other cache nodes. In centrality-based caching [13], the content is stored only once and that is in the cache node which has the highest betweenness centrality. However, the higher betweenness centrality does not ensure that the replicated content is popular in the locality of cache node. This work defines a modified version of LCE [7,10], a replication algorithm called the Conditional LCE (CLCE). In

This work was supported by the Institute for Information & communications Technology Promotion (IITP) grant funded by Korea government (MSIP) (No. R-20160302-003082, Standards development for service control and contents delivery for smart signage services).

Muhammad Bilal is with University of Science and Technology Korea, Electronics and Telecommunications Research Institute (ETRI Campus), 305-700 Daejeon, Rep. of Korea (e-mail: mbilal@etri.re.kr).

Shin-Gak Kang is with Electronics and Telecommunications Research Institute, 305-700 Daejeon, Rep. of Korea (e-mail: sgkang@etri.re.kr).

the CLCE a cache node stores the content if it satisfies a qualifying condition. CLCE also defines a composite function that determines the priority of downloaded content, where the functions are independent of each other and defined and managed by the local administrator to reflect the application specific, operator related control and business model demands. CLCE ensures the quick replication of content in the entire network and ensures that replicated content is popular in the locality of the cache node.

The eviction policy governs the replacement of existing content with arriving content. The eviction policy ensures that the cache node stores the popular content. Several eviction policies have been discussed in the literature [14-24]. In the Independent Reference Model (IRM) the Least Frequently Used (LFU) eviction policy is considered to be the optimum solution [14-15]. However, the LFU has a serious limitation for practical usage; it is impractical to maintain the complete history of the frequency of access for all the stored content. Moreover, the LFU cache performance significantly decreases with burst arrival requests; for instance, the LFU performance is suboptimal for hyper-exponential long-tail traffic models [16]. Consequently, some substitutes were recommended in [18-20]. In [18] aging was introduced to reduce the price of keeping the history of the frequency of access, but the performance of the proposed scheme was considerably low in dynamic and burst request arrival events. In [19] the author presented the Window-LFU; this so-called practical solution for LFU maintains the history of the frequency of access for a limited number of access requests, called a window, but the size of the window is directly proportional to the cache size and the total amount of content in the network. In [20] the author used the LFU algorithm as a filter for sizeable recent history to determine the most frequent item and then to replace it with one of the items in the content of the cache node using any arbitrary eviction policy. The window size used in [20] for the LFU algorithm was the same as that defined by [19] except that a small portion of the cache size was observed.

'In-network caching' is not a new concept; it already existed as an application layer solution for traditional internet architecture. However, cache dynamics in the ICN are different; the cache content in an ICN are subject to change more rapidly, and the cache size is much smaller than dedicated application solutions such as web caching and CDNs. Therefore, the eviction policy should be fast and lightweight. Moreover, an ICN is a network level solution, and the effectiveness of the caching strategy is highly network dependent. To ensure the stability of a cache network, the eviction policy must be ergodic at the node and network levels [25]. Least Recently Used (LRU) [7,21-24] is the most popular eviction policy for ICN architecture; however, LRU does not ensure the storage of popular content. This paper presents an efficient content eviction policy based on an approximation of LFU and a partitioned LRU called the Least Frequent Recently Use (LFRU) eviction policy. In LFRU, the cache is divided into privileged and unprivileged partitions. The privileged partitions use the LRU replacement policy while the unprivileged partition employs an approximated LFU (ALFU). The ALFU keeps the access history of the content for a limited time window ($W$). LFRU evicts the stored content if arriving content meets a qualifying condition; hence, in the company of the LFRU eviction policy, the LCE replication scheme becomes a CLCE replication scheme. The remainder of the paper is organized as follows. In section-II, we render a brief system overview and describe the proposed scheme in detail. Section-III describes the theoretical analysis of LFRU. In Section-IV we compare the theoretical and simulation results and present the performance comparison of LFRU against well-known schemes. Finally, we provide concluding remarks in Section –V.

## II. ANALYTICAL MODEL AND PROPOSED SCHEME

The analytical model used in this study does not follow any particular ICN architecture; instead, it considers each node in the network as a cache node. Let the cache network have a total of $m$ number of cache nodes and $C$ number of content items. A node that generates the content is called a publisher node, and a node that requests content is called a consumer node; a consumer or publisher node can be a human held device or automated machine. The published content should be permanently stored in at least one cache node; it can be a publisher node or any other custodian node. The content is temporarily stored in few intermediate cache nodes (based on the CLCE replication scheme) while it is being delivered to a consumer. If content requests traverse a cache node that holds a temporarily cached copy of that particular content, then the request is entertained locally without being routed towards the publisher.

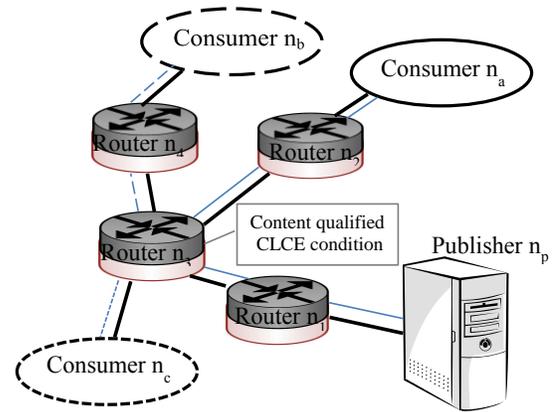

Fig. 1.  A simple Cache Network scenario.

A simple scenario is shown in Figure 1: at time $t$ a consumer $n_a$ requested content published by $n_p$, assuming that intermediate cache $n_3$ qualified the CLCE replication condition; then after successful delivery of the requested content node $n_3$ also holds the copy of content. Now a request for content $c_i$ by consumers $n_b$ and $n_c$ can get the same content $c_i$ from $n_3$ without visiting publisher node $n_p$. The 'in-network caching' significantly improves the content delivery time and decreases the network traffic.

*Notations:*
$C_{ALFU} = C_{ALFU}$ = Set of content in the unprivileged partition
$C_{LRU}^k$ = Set of content in the $kth$ privileged partition

$C_j$ = Set of content in the $j$th cache node

$N_{ALFU}$ = Set of values of counters associated with each cache location in the unprivileged partition $\forall c_i \in C_{ALFU}$

$N_{LRU}$ = Set of values of counters associated with each sub-partition in the privileged partition

$N_{UR}$ = Set of values of counter associated with un-responded requested items. $N_{UR}(i)$ represents the counter associated with un-responded requested items $c_i \notin C_j$.

$c_{min}$ = Content with the lowest counter value in the unprivileged partition

$c_{least}^k$ = Least used content in the $k$th privileged sub-partition.

$v(i)$ = Popularity of content stored at the $i$th location of a cache node

$N = \{n_1, n_2 \dots n_m\}$ = Set of cache nodes

$n_i^u$ = The unprivileged partition

$S_j = \{c_1, c_2, \dots c_n\}$ = The set of content in a cache, it represents the state of the cache at any given instance $j$

$c_i$ = The $i$th rank content in the network

$\lambda_j$ = Request arrival rate at $n_j$

$|n_j|$ = The size of the $j$th cache node in terms of the number of content items that can be stored

$\hat{r}_j$ = The request rate of a newly arrival content at the $j$th cache node

### A. Proposed Scheme

In an ICN kind of cache network, the state of a cache node changes more rapidly, and the cache size is much smaller than dedicated application level solutions such as web caching and CDNs [1-6]; hence the eviction policy should be fast and lightweight. Due to simple and easy implementation, the LRU eviction policy is the most popular eviction policy used for ICN architectures [7,21-24]. However, LRU cannot well conserve the popularity of the content. On the other hand, LFU can conserve the popularity of the content, but LFU has a serious limitation in practical use: It is impractical to maintain the complete history of the frequency of access of all the stored content. Moreover, the performance of the LFU cache decreases significantly with burst request arrivals. This section presents an efficient cache eviction policy based on the approximation of LFU and partitioned LRU called the Least Frequent Recently Use (LFRU) eviction policy; an overview of proposed scheme is shown in Figure 2. LFRU evicts the stored content from the unprivileged partition of $j$th cache if the arrival content request rate is higher than the minimum normalized counter value of the content in the unprivileged partition i-e; $\hat{r}_j(i) \geq min(N_{ALFU})$ and has higher priority compared to victim content i-e; $P(c_i) \geq P(min(N_{ALFU}))$; hence in the company of the LFRU eviction policy the LCE replication scheme becomes a CLCE replication scheme. For practical implementation $\hat{r}_j(i)$ is determined by counter value $N_{UR}(i)$; for theoretical analysis the $\hat{r}_j(i)$ is determined by equation 7.

In LFRU, the cache is divided into privileged and unprivileged partitions as shown in Figure 3. The privileged partitions use the LRU replacement policy while the unprivileged partition employs an approximated LFU (ALFU) scheme. The ALFU keeps the access history of content for a limited time window ($W_T$). The unprivileged partition size should be small enough to cut the operational cost of monitoring the counter value for each memory location and large enough to maximize cache hit probability. The privileged partition is further divided into a $K$ number of sub-partitions. Every privileged partition is monitored by a counter that counts the total number of hits observed in a particular partition within time window $W_T$. The third form of the counter list is associated with un-responded requested items; in CCN architecture this information can be taken from the Pending Interest Table (PIT).

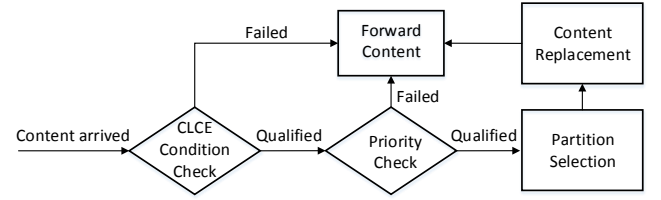

Fig. 2. Overview of LFRU Scheme.

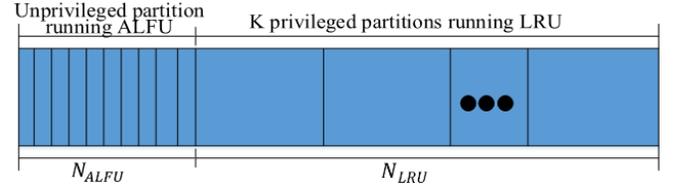

Fig. 3. The Cache Structure Under LFRU Scheme.

For a fixed-size privileged partition, increasing the number of partitions increases the probability of convergence of the most frequently used items within the same partition. In a broader sense, if we consider each privileged partition a box representing one large memory location, then this box is the higher frequency item of ALFU. If $K = |Size\ of\ Privileged\ partition|$ the LFRU scheme becomes ALFU.

*LFRU- Algorithm:*

$P(c_i) = f_0 \circ f_1 \circ f_2 \dots \circ f_k$

$if\ \hat{r}_j(i) \geq min(N_{ALFU}) \wedge P(c_i) \geq P(min(N_{ALFU}))\ for\ c_i \notin$

$C_j$ // CLCE replication condition and priority check

    $if\ min(N_{ALFU}) \leq \hat{r}_j(i) \leq max(N_{ALFU})$

    $C_{ALFU} = (C_{ALFU} - c_{min}) \cup c_i$

    $else\ if\ \hat{r}_j(i) > max(N_{ALFU})$ //Partition selection condition-1

        $\epsilon_o = |N_{LRU}^k - \hat{r}_j(i)|\ \forall\ k = 1,2,\dots K$ //Partition selection condition-2

        insert in sub-partition with smallest $\epsilon_o$

        $C_{ALFU} = (C_{ALFU} - c_{min}) \cup c_{least}^k$

        $C_{LRU}^k = (C_{LRU}^k - c_{least}^k) \cup c_i$

$else$

    Forward the content without storing

LFRU makes the eviction or replacement decision based upon three conditions; a brief stepwise explanation is presented below.

Step 1: Within $W_T$ check two conditions 1) if the request arrival rate for a downloaded content $c_i$ is greater than or equal to the minimum normalized request rate of content in the unprivileged partition (CLCE) and 2) if the content $c_i$ has the priority (determine by the composite function $P$ defined by the local administrator) higher than the replacement candidate content ($min(N_{ALFU})$), and then run LFRU or otherwise forward the downloaded content without storing.

Step 2: If the request arrival rate for a downloaded content item is higher than the minimum normalized request rate and less than the maximum normalized request rate of the content in the unprivileged partition, then evict the minimum normalized request rate content $c_{min}$ from the unprivileged partition and insert downloaded content $c_i$ in the unprivileged partition.

Step 3: If the request arrival rate for a downloaded content item is higher than the maximum normalized request rate of the content in the unprivileged partition, then choose the privileged partition with the smallest $\epsilon_o$ value. Push the least recently used content $c_{least}^k$ from the selected privileged partition into the unprivileged partition and evict the minimum normalized request rate content $c_{min}$ from the unprivileged partition. Finally, insert the downloaded content $c_i$ in the selected privileged partition.

In step 1 the priority of content $c_i$ is defined by a composite function $P(c_i)=f_0{}^\circ f_1{}^\circ f_2 \ldots {}^\circ f_k$. The functions $f_0{}^\circ f_1{}^\circ f_2 \ldots {}^\circ f_k$ are independent of each other and defined and managed by the local administrator to reflect the application specific, operator related control and business model demands. In the CDN the content is cached for the specific content provider with whom the CDN service provider has a commercial relationship; however, we foresee that the content in the ICN will be divided into two broader categories: the neutral content and the content from the publisher with whom the cache owner has the commercial relationship. The caches under commercial contract will give higher priority to the commercial content and/or will provide caching service to neutral content only if the cache has an empty space. However, for the edge ISP network customer satisfaction is of vital importance; here, it is necessary for the caches in the local ISP to be neutral and to define the replication priority function to enhance the customer's quality of experience (QoE) and cache utilization. In either case, in addition to the CLCE replication conditions, the replication scheme is also affected by the locally defined priority function $P$. In this study, we calculated the priority of content based on the arrival rate and size of the content, which is given by

$$P(c_i) = \frac{|c_i|}{n_j} * \frac{\sum N_{ALFU}}{\hat{\tau}_j(i)} \quad (1)$$

### B. Estimating Download Delay

In a cache network, the time elapsed between a content request and availability of content depends on the probability of a hit in cache nodes between the requesting node and publisher or the content custodian node. Let's suppose node $n_i$ requested the content published by node $n_j$; let the maximum delay between $n_i$ and $n_j$ be represented by $d_{ij}^{max}(k)$, where $k$ is the total number of hops between $n_i$ and $n_j$. Node $n_i$ will receive the requested content $c_r$ with the maximum delay $d_{ij}^{max}(k)$ if the intermediate cache nodes do not hold a copy of $c_r$; then the expected delay can be calculated as given below.

$$d_{ij}(k) = \sum_{n=1}^{k} d_{ij}^{max}(k) \, P_h^r(n) \prod_{m=1}^{n}(1 - P_h^r(m)) \quad (2)$$

where $P_h^r(k)$ represents the probability of hit of content $c_r$ at cache node $n_k$.

### C. The Expected Number of Requests in Time Window

For the analysis let us define a zipf-like popularity [26] of content distribution as given below.

$$Z(i) = \frac{1}{\zeta(\alpha)i^{\alpha}} => Z(1) \geq Z(2) \geq \cdots Z(C) \quad (3)$$

$$\zeta(\alpha) = \sum_{i=1}^{C} \frac{1}{i^{\alpha}} \quad (4)$$

where the typical value of $\propto$ is 0.6-1.2, and $\zeta$ is a normalizing constant. Further, suppose $W_T$ is divided into $|W|$ number of small $\delta$ intervals. For a Poisson request arrival with rate $\lambda_j$ the probability of an independent request arrival in interval $\delta$ is given by $\lambda_j\delta$, then the probability that the arrival request in interval $\delta$ is a request for content $c_i$ is provided by

$$P(i,j) = \lambda_j\delta . \frac{1}{\zeta(\alpha)i^{\alpha}} \quad (5)$$

Let us consider $\tau_j$ is an array of random variables where $\tau_j(i)$ is Bernoulli r.v of content $c_i$ and defined as

$$\tau_j(i) = \begin{cases} 1 & with\ P(i,j) \\ 0 & with\ 1 - P(i,j) \end{cases} \quad (6)$$

Further $\hat{\tau}_j$ is an array of random variables that counts the success of $\tau_j$ and is given by

$$\hat{\tau}_j(i) = \sum \tau_j(i) \ \forall \ c_i in\ W. \quad (7)$$

And the expected value of $\hat{\tau}_j$ is given by

$$E\left(\hat{\tau}_j(i)\right) = \frac{|W|}{\zeta(\alpha)i^{\alpha}} \lambda_j \sum_{k=1}^{|W|} \delta_k \quad (8)$$

$$E\left(\hat{\tau}_j(i)\right) = \frac{|W|^2}{\zeta(\alpha)i^{\alpha}} \lambda_j \quad (9)$$

Equation 9 is the expected number of requests in window $W$ for a content $c_i$.

### D. Time Window Size

For web caching a window based LFU scheme (WLFU) was proposed in [19]. The authors proved that a good estimation of the window size could be made based upon the cache size; for instance, the window size for cache $n_j$ is given by

$$|W| = \max\{\theta(|n_j|^3 \ln|n_j| \ln C \ln\frac{1}{\epsilon}), \theta(|n_j| \ln^2 C \ln\frac{1}{\epsilon})\} \quad (10)$$

where $C$ is the total amount of content and $\epsilon > 0$ is any constant. This window estimation is not suitable for our model for the following reasons: 1) unlike WLFU, the LFRU replacement policy is influenced by the arrival rate of un-responded content requests, and it is not necessary that every newly downloaded content item replace the old content; instead an eviction occurs at cache $n_j$ if $\hat{\tau}_j(i) \geq min(N_{ALFU})$. 2) In [19] the window estimation is made to keep track of the number of requests, while LRFU makes an estimation of the appropriate time interval that is sufficient to approximate the history of the cache content. Hence, if the window size is small compared to the expected round trip delay observed by missing items, then it will not be a valuable approximation. 3) The estimated window size in [19] is directly proportional to the cache size and amount of content; maintaining such a large amount of request history is impractical for highly dynamic cache networks.

The probability of a good estimation of the window size with an acceptable error $\epsilon$ and confidence of $Conf\%$ can be calculated using Chebyshev's inequality as given below.

$$P\left[\left|A_w(i) - E\left(\hat{\tau}_j(i)\right)\right| \geq \epsilon\right] \leq \epsilon \leq \frac{\sigma^2}{W\epsilon^2} \leq \left(\frac{100 - Conf}{100}\right) \quad (11)$$

where $A_w(i) = \frac{\sum E\left(\tau_j(i)\right)}{W}$ is the average number of requests for $c_i$ arriving in the last $W$ arrivals, and $c_i$ is the most popular content in $W$. As $\tau_j(i)$ is Bernoulli r.v and $0 < P(i,j) < 1$ then we get

$$\sigma = \sqrt{P(i,j)(1 - p(i,j)} \leq \frac{1}{2} \quad (12)$$

From equations 11 and 12 we can estimate $W$ as

$$W = \frac{100}{4\epsilon^2(100 - Conf)} \quad (13)$$

However, equation 13 is not a tight bound and gives a larger value. For a tighter estimation we use the central limit theorem and standardize the inequality 11 as given below.

$$Z_w = \frac{\hat{\tau}_j(i) - WE(\hat{\tau}_j(i))}{\sigma\sqrt{W}} \quad (14)$$

Now $Z_w$ is a standard normal r.v with $(Z_w) = 0$, and $Var(Z_w) = 1$ and we get

$$P\left[\left|A_w(i) - E\left(\hat{\tau}_j(i)\right)\right| \geq \epsilon\right] \leq \frac{\sigma^2}{W\epsilon^2} \approx P(|Z_w| \geq \frac{\epsilon\sqrt{W}}{\sigma}) \quad (15)$$

Further for $\sigma \leq 1/2$

$$P(|Z_w| \geq \frac{\epsilon}{\sigma}\sqrt{W} \leq P(|Z_w| \geq 2\epsilon\sqrt{W}) \quad (16)$$

$$P(|Z_w| \geq 2\epsilon\sqrt{W}) \leq \left(\frac{100 - Conf}{100}\right) \quad (17)$$

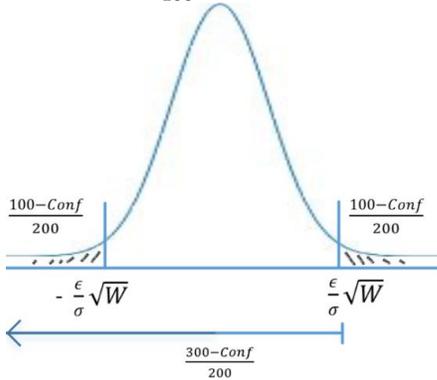

Fig. 4. The PDF of r.v $Z_w$.

The estimated value of $W$ for a given error $\epsilon$ and confidence level $Conf$ can be calculated using the Standard Normal Probabilities table. As shown in Figure 4, let the corresponding value of $\frac{300 - Conf}{200}$ from the Standard Normal Probabilities table be given by $q$ and then the approximation of window $W$ is given by

$$W = \frac{\sigma q}{\epsilon} \quad (18)$$

The estimation is good with an expected error of $\epsilon$ and confidence of $Conf\%$. For a more accurate estimation we define a function $F(W)$:

$$F(W) = E\left(\hat{\tau}_j(i)\right) - A_w(i) - \epsilon \quad (19)$$

In function $F$ only $W$ is unknown, and to determine the value of $W$ we use the 'Newton method'. We use the estimated values from equation 17 as an initial guess and perform a Newton iteration until the value of $W$ converges to a constant value.

$$W_{new} = W_{old} - \frac{F(W)}{F'(W)} \quad (20)$$

where $F'$ is the derivative of function $F$. As discussed earlier the cache content in an ICN changes rapidly, and the replacement policy is also influenced by the arrival rate of un-responded content requests. To make sure $W \geq Avg\left(d_{jk}^l \forall c_l \in C_j\right)$ choose the window observation time as follows:

$$W_T = \max\left\{\frac{W}{\lambda_j}, Avg\left(d_{jk}^l \forall c_l \notin C_j\right)\right\} \quad (21)$$

From equation 2 we know $d_{jk}^l$ is the delay time of $c_l$ requested by $n_j$ and published by $n_k$. The window size specified by equation 21 is small enough to be suitable for practical application and large enough to give a good estimation for popular content. However, if the duration of a burst of a 'content request' dominates the window, then there is a possibility that unpopular content will be pushed into the privileged partition.

## III. THEORETICAL ANALYSIS

### A. Ergodicity of LFRU

A system is said to be stable and converges to a steady state if it is ergodic. The ergodicity of the cache network using LFRU can be established by the following theorems:

*Theorem 1:* The cache with an LFRU policy is individually ergodic.

*Proof of Theorem 1:* Let us consider the cache content at any instance $i$ is represented by cache state $S_i = \{c_1, c_2, \ldots c_k\}$. Any request for content $c_m \forall m = 1,2 \ldots k$ changes the state of the cache from $S_i$ to $S_j$ with the probability $\pi_j$ independent of past transitions. Then vector $S = \{S_1, S_2, \ldots S_n\}$ is a Markov Chain (MC) and each element is a Markov state. Theorem 1 can be proved by proving that $S$ is ergodic. A Markov state is ergodic if it is recurrent and aperiodic. If all states of a MC are single class recurrent and aperiodic, then the chain is said to be ergodic.

*Lemma 1:* The cache state $S_k$ under an LFRU eviction policy is recurrent.

Proof: A state $S_k$ is said to be recurrent if the probability of it returning to state $S_k$ in a finite time $(T_k)$ is 1, i-e $P_k[T_k] = P_k[T_k < \infty | S_0 = S_k] = 1$. Depending upon the popularity of the content in the cache, a change in the state in an LFRU cache requires a variable number of requests; this suggests that the probability of staying in a few states is higher compared to some other states. However, there is always an escape and inward probability for all states, which suggests that the probability of returning to any state in finite time is 1. Hence, the finite LFRU MC is recurrent. It also proves that the LFRU MC is irreducible.

*Lemma 2:* An LFRU MC state $S_k \in S$ is aperiodic.

Proof: The period of a state $S_k \in S$ is given by $b_k = \gcd\{i \geq 1 | P_{kk} > 0\}$. The state $S_k$ is said to be aperiodic if $b_k = 1$, which suggests that any state with self-transition is aperiodic. In LFRU there is always a positive probability of self-transition for all $S_k \in S$; hence, the LFRU MC is aperiodic. Lemmas 1 and 2 conclude the proof of Theorem 1. ∎

*Theorem 2*: A cache with an LFRU policy is individually non-protective.

*Proof of Theorem 2:* An eviction policy is said to be non-protective if all the content in any $jth$ cache at any given time has eviction probability $P_{ev}^j > 0 \, \forall \, c_i \in C_j$. As discussed above the probability of staying in a few states in the LFRU MC is higher compared to other states, but there is always an escape probability. Let us suppose that $c_h$ is highly popular content and $c_h \in C_j$; if content $c_h$ is in an unprotected partition, it will take longer to evict any content from the protected partition. Without changing the content if $c_h$ is pushed into the protected partition, then the time required to evict any content from the protected partition will decrease. In either case, there is invariably a positive probability of eviction of any content; hence the LFRU eviction policy is non-protective. ∎

*Theorem 3*: A feed forward cache network with an LFRU cache policy is an ergodic system.

*Proof of Theorem 3:* According to theorem 4, which is discussed in [25], a CN is proved to be ergodic if its caches are individually ergodic and non-protective. As we have shown that LFRU is an individually ergodic and non-protective eviction scheme, a cache network of LFRU is also ergodic. ∎

### B. Cache Hit Rate

The LFRU is composed of approximated LFU and protected LRU policies; thus, the cache hit rate is the sum of the hit rate of approximated LFU and protected LRU policies.

$$h_{LFRU} = h_u + \sum_{i=1}^{k} h_p^i \quad (22)$$

where $h_u$ is the cache hit rate in the unprivileged partition with the ALFU scheme, $h_p^i$ is the cache hit rate in the $ith$ privileged partition with the LRU scheme, and $k$ is the total number of privileged partitions.

#### 1) Cache Hit Rate in unprivileged partition $(h_u)$

The ALFU is an approximation of the LFU scheme, and the accuracy of approximation is determined by the appropriate choice of window size. Considering the LFU eviction policy and zipf law with $\alpha = 1$, the cache hit rate can be determined by $h_{LFU} = \sum_{i \in c_{ALFU}} \frac{1}{i \ln c}$, and the cache hit rate for ALFU is given by

$$h_u = \sum_{i \in c_{ALFU}} \frac{1}{i \ln c} \text{x} \ (1 - P[W \text{ is weak estimator}]) \quad (23)$$

$$h_u \geq (1 - \epsilon) \sum_{i \in c_{ALFU}} \frac{1}{i \ln c} \quad (24)$$

where $\epsilon$ is an estimated error for a window $|W|$.

#### 2) Cache Hit Rate in privileged partition $(h_p^i)$

The probability of a hit in the privileged partition is the sum of the probabilities of a hit in sub partitions. Let us suppose $|n_j^k|$ is the size of the $kth$ partition at the $jth$ cache node, $X(\tau_n)$ is the total number of requests arriving at cache $j$ and $X_c(\tau_n)$ is the number of requests arriving with sub-partition selection conditions 1 and 2 (refer to the algorithm) within a time interval $(0, \tau_n)$. $v(n)$ is the popularity of the $nth$ content and $T_j(n)$ is the time at which precisely $|n_j^k|$ number of requests other than $c_n$ arrive with selection conditions 1 and 2. Let $c_n$ be downloaded to the $kth$ partition at the $jth$ cache node, and the next request for content $c_n$ will be hit at time $\tau_n$ if $X_c(\tau_n) < |n_j^k|$. Note that $X(\tau_n) \geq |n_j^k|$. Then,

$$\{X_c(\tau_n) < |n_j^k|\} = \{T_j(n) > \tau_n\} \quad (25)$$

Now, assuming Poisson arrival of requests, the probability of a hit for $c_n$ in the $kth$ partition at the $jth$ cache node is given by

$$P_h^n = P(T_j(n) > \tau_n) = E(1 - e^{-v(n)T_j(n)}) \quad (26)$$

According to Che's approximation [27], $T_j(n)$ is a constant independent of $n$ i-e, $T_j(n) = T_j$ . Consequently, we get

$$|n_j^k| = \text{Const} = \sum_{n \in N}(1 - e^{-v(n)T_j}) \quad (27)$$

where $T_j$ is the singular root of equation 27; Che et al. referred to it as the "characteristic cache time". From equations 25 and 26 we get the hit rate for content $c_n$ for $1 \leq n \leq N$:

$$P_h^n = 1 - e^{-v(n)T_j} \quad (28)$$

$T_j$ is the Che approximation of time at which exactly $|n_j^k|$ number of requests other than $c_n$ arrive with selection conditions 1 and 2; however, during time $T_j$ there will be a large number of other requests also arriving that do not fulfill the selection conditions 1 and 2. That suggests that the 'characteristic cache time' without selection conditions 1 and 2 $T_j'$ is much smaller than $T_j$. In equation 27 $T_j$ is unknown. To determine the value of $T_j$, equation 27 can be solved using Newton's method. Let us define the function $f(T_j)$:

$$f(T_j) = \text{Const} - \sum_{n \in N}(1 - e^{-v(n)T_j}) \quad (29)$$

Let the following intuitive equation be the initial guess for a Newton iteration:

$$T = \frac{\text{Const}}{\sum_{n \in N} v(n)} \quad (30)$$

After the initial guess we perform a Newton iteration (given in equation 31) until the value of $T$ converges to some constant value after a few iterations:

$$T_j^{new} = T_j^{old} - \frac{f(T_j^{old})}{f'(T_j^{old})} \quad (31)$$

where $f'(T_j)$ is the derivative of function $f(T_j)$ defined in equation 29.

Now all the parameters in equation 24 are known; hence the cache hit rate for the privileged sub-partition $n_i$ is given by

$$h_p^i = \frac{1}{|n_j^k|} \sum_{n \in C_{LRU}^k} P_h^n \quad (32)$$

Finally, from equations 22, 24, 26 and 32, the cache hit rate for LFRU with $k$ number of privilege partitions is given by

$$h_{LFRU} \geq (1 - \epsilon) \sum_{n \in C_{ALFU}} \frac{1}{n \ln c} + \sum_{j=1}^{k} \sum_{n \in C_{LRU}^k} P_h^n \ (1 - e^{-v(n)T_j}) \quad (33)$$

where $\epsilon$ is an estimated error for a window $|W|$.

## IV. THE VERIFICATION OF THEORETICAL ANALYSIS AND SIMULATION RESULTS

### A. The Network Setup

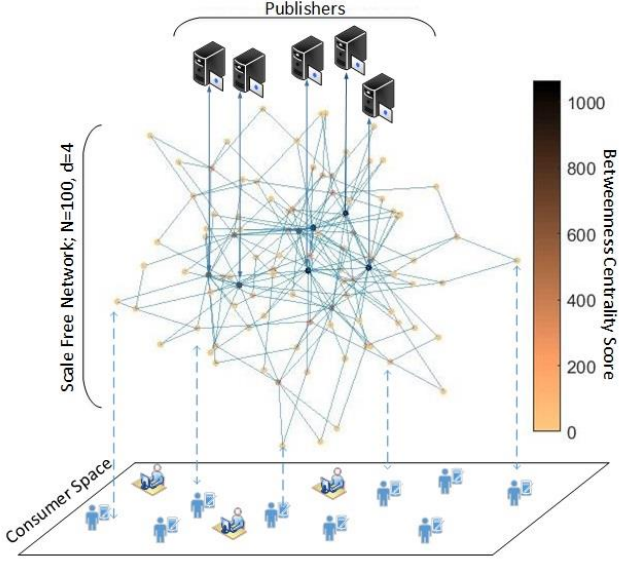

Fig. 5. A depiction of network setup used for verification of theoretical analysis and simulation.

Let us consider a scale-free network of 100 cache nodes generated using the Barabási–Albert (BA) model, as shown in Figure 5, which connects the publisher and the consumer space. Each cache node has a static request routing table. Further, assume that there are 5 content publishers in the network, each with 10000 content itmes; a Zipf-distribution as given in equation 2 determines the population of 50000 content items in the entire network. The content publishers are connected to the cache nodes with a top betweenness centrality score; in Figure 5 the darker the colors of the dot the higher the betweenness centrality score of the cache node. Furthermore, assume that each cache node is also connected to a large number of consumers; in Figure 5, the dotted line represents the aggregated content request arrival at a cache node directly from the consumer space. The total request arrival rate ($\lambda_j$) at any given cache node $j$ is the sum of the aggregated content requests directly coming from the consumer space ($\lambda_j^d$) plus aggregate requests forwarded by the neighboring cache nodes ($\lambda_j^f = \sum \lambda_{i,j}$, where $i \in set\ of\ adjacent\ cache\ nodes$).

### B. The Verification of Theoretical Analysis

In Sections III-A and B we proved that a cache network converges to a steady state if it uses the LFRU scheme. Further, we also estimated the probability of a hit for LFRU. In this section, we verify the theoretical analysis by comparing the theoretical convergence and probability of a hit with the simulation results. Let's suppose the sequence of random variables $v(i,j,t), i = 1,2,\dots |n_j|$, representing the popularity of the content stored at cache node $n_j$ at any instance $t$. All $v(i,j,t), i = 1,2,\dots |n_j|$ having the same finite mean ($\mu = \frac{\zeta(\alpha-1)}{\zeta(\alpha)}$) and variance ($\sigma = \frac{\zeta(\alpha-2)}{\zeta(\alpha)} - \left(\frac{\zeta(\alpha-1)}{\zeta(\alpha)}\right)^2$), then the average of $v(i,j,t), i = 1,2,\dots |n_j|$ is given by

$$X_n(j,t) = \frac{1}{|n_j|} \sum_{i=1}^{|n_j|} v(i,j,t) \qquad (34)$$

Our proof strategy is based on the assessment that when the whole network approaches a steady state the ratio $\frac{X_n(j,t+1)}{X_n(j,t)}$ approaches 1. In other words, the state of the cache node becomes persistent. It also implies that $\lambda_j^f$ (aggregate requests forwarded by the neighboring cache nodes) at each cache node also converges. The total request arrival rate based on theoretical calculations is determined as follows:

1) Let us assume that the request arrival process at any cache node $j$ is a Poisson with a rate of $\lambda_j = [700, 1000]$ and each cache node can store 3000 items of content.

2) Apply the routing table and determine the outgoing request rates at each cache node.

3) Update the state of all cache nodes based on the probability of a hit of a content LFRU cache node using equation 34.

4) Calculate the request arrival rate at each cache node, which is the sum of the aggregated content requests from the consumer space generated with a random Poisson process plus the requests forwarded by the neighboring cache nodes, determined at step 2.

5) Check the steady state condition $\frac{X_n(j,t)}{X_n(j,t)} \geq 1 - \epsilon$ for each cache node, where $\epsilon$=0.001 is a small number.

6) Repeat steps 2 to 5 until the steady state condition becomes true. The value of $\epsilon$ is introduced to reduce the number of repetitions from steps 2 to 5.

7) Finally, report the $\lambda_j^f$ for each cache node in a steady state.

Once the average request rate forwarded by the cache node in a steady state is available, we compare it with the corresponding results obtained by simulating a similar network setup in Matlab. The Q-Q plot in Figure 6 shows that both the theoretical and simulation implementations converge to a similar form of steady state with a slight deviation. The Q-Q plot indicates that under a steady state the aggregate forwarded requests ($\lambda_j^f$) received by the majority of the cache nodes is slightly higher in the simulation than in the theoretical estimation. The degradation in the performance of the simulation is due to the fact that the independent Poisson request arrival does not hold because the request forwarded by the intermediated cache nodes are not the independent Poisson process.

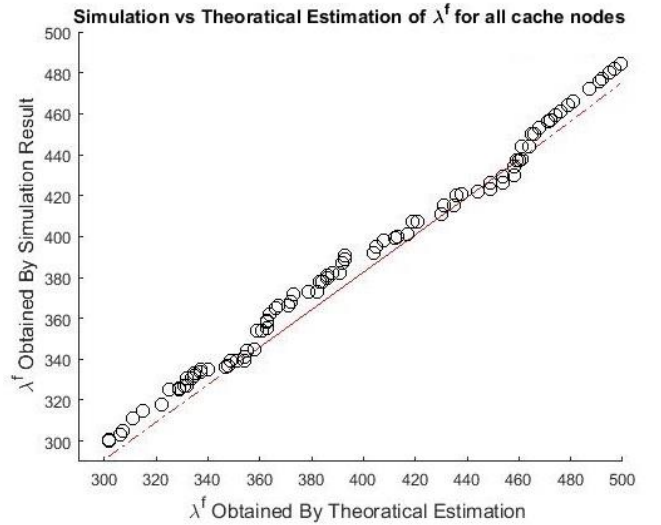

Fig. 6. Q-Q plot for the comparison of requests forwarded by each cache nodes in steady state; obtained by theoretical estimation and simulation.

For simplicity, in the above analysis, we considered the simple priority function as described in equation 1. In a future study, it will be interesting to investigate a similar analysis when different cache nodes define different priority functions. Moreover, if the cache size $|n_j| \to \infty$ then according to the weak law of large numbers and the central limit theorem equation 34 converges to standard normal distribution, which can deduce further interesting findings; for instance, convergence in probability can help in estimating the popularity of content in the cache node's locality. However, in this study, we consider a cache with a limited size and leave the further analysis of an infinite cache for future work.

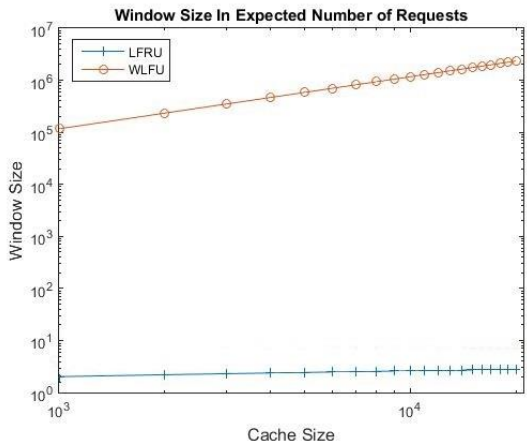

Fig. 7 Windows Size comparison between LRFU and WLFU for different cache size

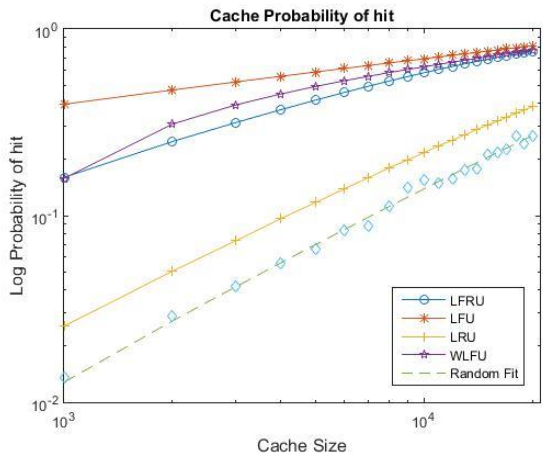

Fig. 8 The Probability of hit comparison between LFRU, LFU, LRU WLFU and Random for different cache size.

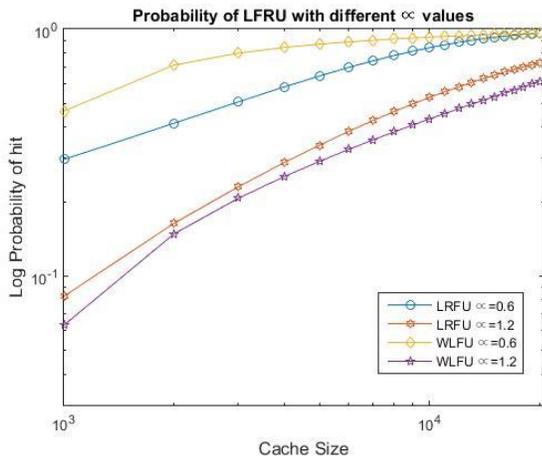

Fig. 9 The Probability of hit comparison between LFRU and WLFU for different cache size and α values.

### C. Performance Comparison

Let us consider the network setup as described above. Further, assume that for the LFRU scheme 20% of the cache is allocated for the unprivileged partition. We implemented the LFRU along with LRU, LFU, Random and WLFU content eviction schemes in Matlab. The results are taken for different cache sizes, as shown in Figures 7 to 9.

The estimated window size (in number of requests) of WLFU and LFRU is shown in Figure 7. The WLFU requires that the history be maintained for a large number of incoming requests; additionally, the window size further increases with an increase in the cache size. Sustaining history for such a large number of request arrivals requires many resources; hence, WLFU is not a practical solution for ICN networks. Figure 8 shows the probability of a hit comparison between LFRU, LFU, WLFU, LRU, and Random eviction schemes; and the results are taken for ∝=0.8. The LFRU outperforms the Random and LRU eviction schemes, while the hit rate is close to the WLFU and LFU schemes though Figure 9 shows that the LRFU probability of a hit increased over WLFU at ∝=1.2. The reason behind this behavior is understandable: In a Zipf-distribution with a higher ∝ value the popularity of content shifts towards the tail and skews towards a higher rank with a lower value of ∝. Hence, with a higher value of ∝, within the unprivileged partition of LFRU the probability to fulfill the CLCE replication condition increases. Thus the chance to force popular content into the privilege partition also increases. Therefore, LFRU performs better than WLFU for a higher value of ∝.

## V. CONCLUSION

In this paper, we proposed cache replication and eviction schemes, Conditional Leave a Copy Everywhere (CLCE) and Least Frequent Recently Used (LFRU), which are well suited for an ICN type of cache network. For the content spread, we used the CLCE replication scheme. The CLCE replication scheme ensures that the content is replicated in the cache node where the content is locally and recently popular. We also introduced the concept of replication priority, which is defined and managed by the local administrator to reflect the application specific, operator related control and business model demands. The LFRU eviction scheme is an approximation of the Least Frequently Used (LFU) scheme coupled with the monitored Least Recently Used (LRU) scheme. LFRU is practically implementable for rapidly changing cache networks like ICNs. We proved that a cache network with LFRU is ergodic and convergent. We also estimated the probability of a hit for LFRU, and finally we compared the performance of LFRU against Random eviction, LFU, LRU and WLFU schemes.

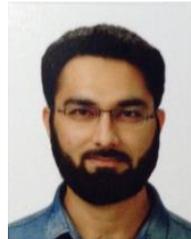

**Muhammad Bilal** has received his BS degree in computer systems engineering from University of Engineering and Technology, Peshawar, Pakistan and MS in computer engineering from Chosun University, Gwangju, Rep. of Korea. Currently, he is Ph.D. student at University of Science and Technology, Korea at Electronics and Telecommunication Research Institute Campus, Daejeon, Rep. of Korea. . He has served as a reviewer of various international Journals including IEEE Communications Letters, IEEE Access, International Journal of Communication Systems and Journal of Network and Computer Applications. He has also served as a program committee member on many international conferences. His primary research interests are Design and Analysis of Network Protocols, Network Architecture, and Future Internet.

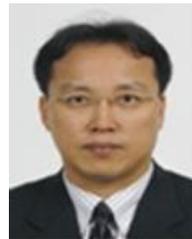

**Shin-Gak Kang** received his BS and MS degree in electronics engineering from Chungnam National University, Rep. of Korea, in 1984 and 1987, respectively and his Ph.D. degree in information communication engineering from Chungnam National University, Rep of Korea in 1998. Since 1984, he is working with Electronics and Telecommunications Research Institute, Daejeon, Rep. of Korea, where he is a principal researcher of infrastructure standard research section. From 2008 he is a professor at the Department of Information and Communication Network Technology, University of Science and Technology, Korea. He is actively participating in various international standard bodies as a Vice-chairman of ITU-T SG11, Convenor of JTC 1/SC 6/WG 7, etc. His research interests include multimedia communications and Applications, ICT converged services, contents networking, and Future Network.